\documentclass[amsmath,amssymb,aps,onecolumn]{revtex4-2}

\usepackage{graphicx}
\usepackage{subcaption}
\usepackage{dcolumn}
\usepackage{bm}

\usepackage{braket}
\usepackage{xcolor}
\usepackage{mathtools}

\usepackage[colorlinks=true,citecolor=blue,linkcolor=blue,urlcolor=blue]{hyperref}

\begin{document}


\title{Probeless vs Probe-Based Variable-Strength Eavesdropping in Quantum Key Distribution}

\author{Yuval Idan}
\affiliation{Faculty of Engineering and the Institute of Nanotechnology and Advanced Materials, Bar-Ilan University, 5290002 Ramat Gan, Israel}
\author{Tal Gofman}
\affiliation{NVIDIA, Hakidma 26, Ofer Industrial Park, Yokneam 2069203, Israel}
\author{Ziv Abelson}
\affiliation{NVIDIA, Hakidma 26, Ofer Industrial Park, Yokneam 2069203, Israel}
\author{Isabelle Cestier}
\affiliation{NVIDIA, Hakidma 26, Ofer Industrial Park, Yokneam 2069203, Israel}
\author{Elad Mentovich}
\affiliation{NVIDIA, Hakidma 26, Ofer Industrial Park, Yokneam 2069203, Israel}
\author{Eliahu Cohen}
\affiliation{Faculty of Engineering and the Institute of Nanotechnology and Advanced Materials, Bar-Ilan University, 5290002 Ramat Gan, Israel.}

\date{\today}

\begin{abstract}
Quantum key distribution (QKD) is a provably secure way of generating a secret key, which can later be used for encoding and decoding information. In this paper we analyze the effects of an eavesdropper's variable-strength measurements on QKD. Two types of measurements have been considered: (i) a probe-based model, commonly referred to as a ``weak measurement'', in which each qubit is weakly coupled to a continuous variable probe which is later projectively measured (ii) a probeless model, usually referred to as a ``partial measurement'', where only a small (tunable) part of all transmitted photons is projectively measured and the rest are transmitted with no disturbance. The information gain of the eavesdropper and the quantum-bit-error-rate (QBER) are computed for each case. 
An experimental realization of an intercept-and-resend attack based on variable-strength partial measurements is demonstrated in a time-bin-encoded, fiber-based simplified Bennett-Brassard 1984 (BB84) protocol, which is compatible with data centers.
It is shown that the measured information gain and QBER follow the theoretical curves across the full coupling range, validating the partial-measurement model and clarifying its relation to the well-known monitoring channel. Further attacks involving photon number splitting and noise injection during the calibration stage are also analyzed.
The results highlight the theoretical differences between weak and partial measurements, while also demonstrating the practicality of probeless eavesdropping in the case of real-world QKD systems. 

\end{abstract}

\maketitle

\section{Introduction}
Quantum key distribution (QKD)~\cite{bennett2020quantum,shor2000simple,pirandola2020advances}
leverages quantum mechanics to generate and share cryptographic keys while
making any eavesdropping attempt detectable.  Recent work has increased key
rates, extended transmission distances, and integrated QKD with conventional
telecommunication networks~\cite{zapatero2023advances,pereira2023modified,
gleim2016secure}.  Although the formal security proofs remain intact%
~\cite{shor2000simple,pirandola2020advances}, several attack strategies exploit
imperfections in Alice’s and Bob’s devices~\cite{lydersen2010hacking,
garcia2020attacking,ashkenazy2024photon}.

Weak measurements~\cite{aharonov1988result,tamir2013introduction} may offer one such avenue: an ancillary probe interacts only weakly with the signal, and therefore projective measurement of the probe leads to slight decoherence of the signal while providing a small amount of information. This potentially provides the eavesdropper with increased flexibility. However, implementing a genuine weak measurement within a QKD scenario could be practically challenging, as another quantum system or another degree of freedom needs to be invoked. A more feasible
alternative is given by partial (probeless) measurements \cite{elitzur2001nonlocal}, where the adversary
measures only a chosen fraction of the transmitted photons, leaving the remainder undisturbed. As shown below, both approaches respect the fundamental information–disturbance trade-off, yet they produce different quantitative signatures in the observed quantum-bit-error rate (QBER). Our work emphasizes the similarities and differences between these two measurement approaches.

Our QKD implementation at NVIDIA Laboratories  is performed as part of an attempt to use short-distance QKD within data centers, where the detector's dead time is the main factor limiting the key rate. Similarly, our realization is based on a practical source of weak coherent states with time-bin-encoding.  In this setting, variable-strength partial measurements were applied to the simplified BB84 protocol~\cite{rusca2018security,boaron2020long}. 
We show that Alice and Bob's QBER and Eve's information gain match the theoretical predictions.
The remainder of the paper is structured as follows. In Sec.~\ref{sec2} the
theoretical background regarding implementations of QKD protocols, as well as on variable-strength measurements with and without a probe are presented. Sec.~\ref{sec: EVE} connects weak measurement protocols to eavesdropping on QKD protocols, emphasizing the information gain and disturbance trade-off, while in Sec.~\ref{sec:pointer} we examine how the particular shape of Eve's pointer affects her performance. In Sec.~\ref{PNS and Weak}, an enhanced attack combining photon number splitting (PNS) and weak measurements is analyzed.
Finally, Sec.~\ref{sec:experiment} presents an experimental demonstration: an eavesdropping attack on a QKD protocol in a data-center setting, and Sec.~\ref{sec:noise injection} augments it with an experimental noise-injection attack on Alice and Bob’s monitoring apparatus during calibration. Conclusions are presented in Sec.~\ref{sec:conco}.

\section{Theoretical background} \label{sec2}

\subsection{Time-bin BB84}
There are various ways to implement QKD protocols with mutually unbiased bases, such as polarization-based QKD~\cite{lo2014secure,bennett1992experimental} and time-bin QKD~ \cite{scarani2009security, rusca2018security}. In this work, we focus on a time-bin-encoded BB84 protocol with weak coherent states. In time-bin BB84 protocols, Alice encodes photons either in the phase basis or the time basis. In the time basis, Alice can encode her information in the early mode denoted as $\ket{e}$ and in the late mode denoted as $\ket{l}$, where $\braket{l|e} = 0$. In the phase basis, Alice may encode the $0$ temporal phase via $\ket{+} = \frac{1}{\sqrt{2}}\left(\ket{e} + \ket{l}\right)$, and the $\pi$ temporal phase via $\ket{-} = \frac{1}{\sqrt{2}} \left(\ket{e}-\ket{l}\right)$, with $\braket{+|-} = 0$ and $|\braket{i|j}|^2= \frac{1}{2}, i \in\{e,l\}, j \in \{+,-\}$.
Bob chooses between time-of-arrival measurement in the early/late basis or phase measurement in the +/- basis, usually performed using an asymmetric Mach-Zehnder interferometer (MZI). Our practical implementation is based on \cite{rusca2018security}, which is known as the ``simplified BB84 protocol''. In this protocol, the secret key is generated through measurements in the time basis while the phase basis is used solely for monitoring Alice and Bob's QBER. This is taken into account when analyzing the experiment, e.g. Eve's information gain can theoretically reach $100\% $, because her measurements are always performed in the time basis where information is encoded. 

\subsubsection{Weak coherent states}
At present, the limited reliability of high-rate single-photon sources hinders their deployment in high-speed QKD systems. Therefore, Alice and Bob use weak coherent states (WCS) to mimic a single-photon source. A coherent state is defined by
\begin{equation}\label{eq:WCS}
    \ket{\alpha}
= e^{-|\alpha|^2/2}
\sum_{n=0}^{\infty}
\frac{\alpha^{n}}{\sqrt{n!}}\,
\ket{n}.    
\end{equation}

Here, \(\{\ket{n}\}_{n=0}^\infty\) are Fock states, and the expectation value of the photon-number operator (with $\hat N=\hat{a}^\dagger \hat{a}$) in the coherent state $|\alpha\rangle$ is
\[
\langle N\rangle
= \bra{\alpha}\hat N\ket{\alpha}
= |\alpha|^2 \equiv \mu,
\]
which represents the mean photon number per pulse of the coherent state. Alice and Bob aim to use the minimum $\mu$ that still maintains a secure, high-speed QKD protocol. A drawback of using a very small $\mu$ is that Alice will emit many vacuum states, thereby slowing down the key generation rate. However, a large $\mu$ may lead to the so-called photon number splitting (PNS) attack \cite{huttner1995quantum,ashkenazy2024photon}, which allows an eavesdropper to split multi-photon pulses and gain additional information during the basis reconciliation step of the protocol. This attack has been extensively studied, and today various protocols and countermeasures exist to protect Alice and Bob from such attacks \cite{scarani2004quantum,branciard2005security,hwang2003quantum, lo2005decoy,stucki2005fast}.
The experiment employs WCSs with mean photon number $\mu \ll 1$. Hence, detection events were mostly single-photon events: 
\begin{equation}
   \ket{\sqrt{\mu}} = \mathcal{N}\left( \ket{0} + \sqrt{\mu}\ket{1} +O(\mu)\right), 
\end{equation}
where $\mathcal{N}$ is a normalization constant.
In practice, there were many null events resulting from the vacuum's significant contribution, but we post-selected on non-null detections. More information about our experimental realization  provided in Sec.~\ref{sec:experiment}.

\subsection{Weak measurements}\label{Sec_weak_measuremet}
A POVM (positive operator‐valued measure)~\cite{nielsen2010quantum,peres2002quantum} represents the most general form of quantum measurement. In practice, they are often implemented by coupling the quantum state to an ancilla through an interaction, and then measuring the ancilla to reveal the properties of the original quantum state in the chosen basis. Although being an important, special case, projective measurements are described by projective-valued measures (PVMs) since they are implemented using a set of projection operators $\hat{P}_i$ which are idempotent and mutually orthogonal, whose sum is the identity. This ensures that each measurement outcome corresponds to a unique state of the system belonging to an orthogonal basis. Both PVMs and the more general POVMs can be implemented using von Neumann's interaction Hamiltonian \cite{tamir2013introduction}. The ratio between the coupling strength and the ancilla's uncertainty determines whether the measurement is projective or {\it weak}. Weak measurements introduce smaller disturbances to the measured states, but they also obtain a smaller amount of information.
Weak measurements can be utilized by an eavesdropper to gain partial information about an arbitrary quantum state. The weak measurement procedure can be derived from the von Neumann measurement model, where one couples the measured state to a measuring pointer (probe), typically having a continuous probability density function (PDF), e.g. a Gaussian.
The explicit form of the mathematical expression of the weak measurement protocol is defined as follows.
For a quantum state $\ket{\psi}$ and a pointer system $\ket{\phi}$, a von Neumann measurement is defined as~\cite{mello2014neumann}: \begin{equation} \ket{\psi}\ket{\phi}\rightarrow e^{-\frac{i\epsilon\hat{A}\otimes \hat{B}}{\hbar}}\ket{\psi}\ket{\phi} = \ket{\tilde{\psi}}\ket{\tilde{\phi}}, \end{equation} where 
$\epsilon$  is the coupling strength~\cite{tamir2013introduction}. The state of the ancillary system changes according to the principal system's quantum state in the $\hat{A}$ basis. Although projective measurements are described by PVMs, weak measurements are classified as POVMs~\cite{tamir2013introduction}. In weak measurements, the interaction is weak meaning that the principal system and the pointer weakly affect each other, leading to minor correlation between them. 
For some arbitrary quantum state $\ket{\psi} = \sum_i\alpha_i\ket{i}$ and a continuous pointer $\ket{\phi} = \int_xf(x)\ket{x}dx$, where $ \int_x|f(x)|^2dx =1$, performing a weak measurement leads to:
\begin{equation}
    e^{-i\frac{\epsilon\hat{A}\otimes \hat{P}_d }{\hbar}}\ket{\psi}\ket{\phi} = \sum_i \alpha_i\ket{i}\int_xf(x-\lambda_i\epsilon )\ket{x}dx,
\end{equation}
where $\hat{P}_d$ is the momentum operator associated with the pointer,  $\lambda_i$ is the eigenvalue of the eigenvector $\ket{i}$ and
$\epsilon \ll 1$ is a tunable parameter that Eve may use to adjust her measurement strength and control the information gain-disturbance trade-off~\cite{fuchs1997optimal}. In this work, we will calculate Alice and Bob's QBER, as well as Eve's information gain as functions of $\epsilon$.  The quantum channel of the weak measurement exerted on the measured particle is defined as:
\begin{equation} \label{eq:channel}
   \mathcal{M}(\rho) =  Tr_d[\hat{U}(\rho\otimes \rho_d)\hat{U}^{\dagger}],
\end{equation}
where $\rho$ is the system's density matrix, $\rho_d$ is the pointer's density matrix in position space and $U$ is the unitary operator that acts on the joint system. In our work, Eve's PDF is a Gaussian defined as: $f(x) = (2\pi\Delta^2)^{-\frac{1}{4}}e^{\frac{-x^2}{4\Delta^2}}$, in the weak measurement limit $\frac{\epsilon}{\Delta}\ll1$. Since the Gaussian is continuous, it helps to approximate Eq.~\eqref{eq:channel} as a monitoring channel~\cite{dieguez2018information}:
\begin{equation} \label{eq: monitoring}
    \mathcal{M}(\rho) = (1-\delta)\rho + \delta \sum_i\hat{A}_i\rho \hat{A}_i ^\dagger ,
\end{equation}
where $\sum_i \hat{A}_i^\dagger \hat{A}_i=\hat I$. For a pointer of  width $\Delta$ and weak coupling obeying $\frac{\epsilon}{\Delta}\ll 1$, tracing out the pointer yields, to leading order, a completely positive trace-preserving (CPTP) map of the form of Eq.~\eqref{eq: monitoring} with a mixing parameter $\delta=O\left(\frac{\epsilon^2}{\Delta^2}\right)$. For BB84 projectors $\{\hat{P}_i\}_{i=1}^4=\{|0\rangle\!\langle 0|,\,|1\rangle\!\langle 1|,\, (|0\rangle{+}|1\rangle)(\langle 0|{+}\langle 1|)/2,\,(|0\rangle{-}|1\rangle)(\langle 0|{-}\langle 1|)/2\}$, choosing $\hat{A}_i=\hat{P}_i/\sqrt{2}$ makes $\sum_i \hat{A}_i^\dagger \hat{A}_i=\hat I$ explicit. In addition to using Eq.~\eqref{eq: monitoring}, we also simulate the exact weak-measurement dynamics (no monitoring approximation), showing that the QBER scales quadratically in $\frac{\epsilon}{\Delta}\ll1$ (and hence in $\delta$).


\subsection{Partial measurement}
Partial measurements, also referred to as probeless measurements are standard projective measurements applied probabilistically to only a subset of signals. Eavesdroppers may realize such an attack when they are unable to carry out the measurement with a probe. Additionally, this method has been studied using the two-state vector formalism (TSVF) and bears interesting relations with erasure and weak values \cite{elitzur2001nonlocal,elitzur2011retrocausal,okamoto2023experimentally}.
Partial measurements are PVMs that an attacker applies to Alice and Bob's system on only a portion of the transmitted photons. In some sense, it is an alternative to the intercept-and-resend attack.
The quantum channel describing partial measurements can be defined, similarly to a monitoring transformation, as:
\begin{equation}
    P(\rho) \equiv \rho' =(1-\gamma)\rho + \sum_{i}\frac{\gamma}{2}\hat{P}_i\rho\hat{P}_i.
\end{equation}
where the projections are the aforementioned $\{\hat{P}_i\}_{i=1}^4$. The channel defined above is a special case of the monitoring channel in Eq.~\eqref{eq: monitoring}, obtained by choosing $\delta = \gamma$. Therefore, we shall henceforth introduce a single parameter
$
  \epsilon \in [0,1],
$
to quantify the measurement–interaction strength, setting $\epsilon = \gamma = \delta$.  In the limit $\epsilon = 1$, the map reduces to a projective (strong) measurement, fully collapsing the wavefunction onto an eigenstate of the measured operator, whereas $\epsilon = 0$ corresponds to no interaction with the measurement device.
The projectors denoted above correspond to the time-bin-encoding used in our QKD protocol. Bob implements the same measurement as Eve, but the state is already disturbed after undergoing this channel, leading to an increase in the QBER.
This quantum channel is similar to the aforementioned weak measurement channel. However, in probe-based weak measurements, the disturbance (QBER) and information gain are also affected by the pointer's wavefunction. When using weak measurements, the pointer is commonly taken to be continuous and symmetric, whereas in partial measurements there is no continuous pointer; outcomes are discrete projective results. We will show that the measurement strength in each method leads to a slightly different trade-off between information gain and QBER. The following sections compare these methods and present supporting experimental results in which Eve employs partial measurements to attack a simplified time-bin BB84 protocol.

\subsection{QBER and information gain}\label{sec:QaG}
The two main figures of merit which are commonly used to evaluate QKD protocols are Alice and Bob's QBER ($Q$) and Eve's information gain ($G$). These two quantifiers correspond to Eve's induced disturbance and Eve's success rate, respectively.
The interaction parameter $\epsilon$ directly affects the shared density matrix $\rho_A$ and correlates with $Q$ and $G$. 
The mathematical definition of $Q$ is the Hamming distance between Alice and Bob's classical bit string of the sifted key:
\begin{equation}
     Q=d_{AB}=\frac{\sum S^A_i\oplus S^{B}_i}{L},
\end{equation}
where $S^{A}$ is Alice's string and $S^{A}_i$ is the value of Alice's $i$-th bit and $L$ is the length of the string.
In the standard intercept-and-resend attack ($\epsilon=1$), Alice and Bob's QBER reaches a maximum of $25\%$ ($Q=0.25$), and on the other hand, if Eve does not interact with Alice's quantum state $\rho_A$ ($\epsilon =\delta =  0$) $Q=0$, thus $0\leq Q\leq 0.25$.
If a third party (i.e. external noise) also interacts with the transmitted qubits the bound may become $0\leq Q\leq 0.5$, where in the case of $Q = 0.5$ Alice's information is completely erased.
Eve's information gain is the number of correct classical bits that she measured relative to Alice's classical string:
\begin{equation}
    G = 1 -\frac{\sum_iS_i^A\oplus S_i^E}{L}.
\end{equation}
The information gain has a direct relation to the mutual information between Alice and Eve \cite{ashkenazy2024photon}.
In the case of variable-strength (weak) measurements, Q and G become continuous rather than discrete functions. They increase monotonically with the measurement strength and depend on the selected configuration of Eve’s measurement device. When the properties of the device are neglected, $Q$ and $G$ follow the theoretical and combinatorial predictions discussed in Sec.~\ref{sec:partial}.

\section{Variable-Strength Eavesdropping}\label{sec: EVE}

\subsection{Weak measurement attack}\label{Sec:weak_measurement_attack}
\begin{figure}
    \centering
    \includegraphics[scale = 0.35]{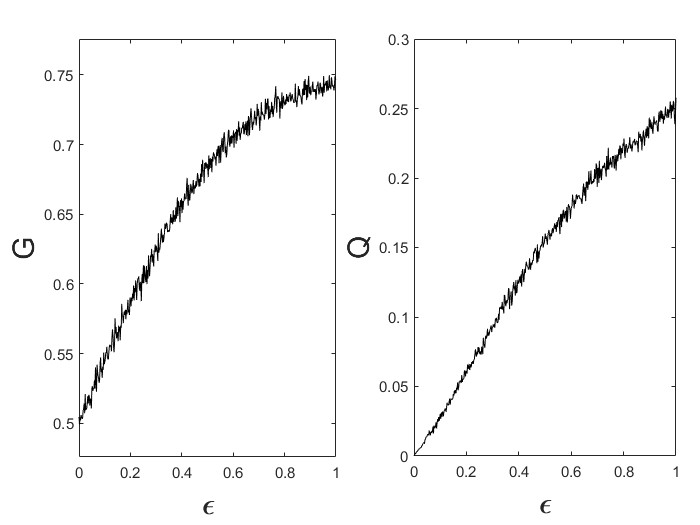} 
    \caption{Information gain (G) and disturbance (Q) for variable-strength probe-based measurements. Left: Eve's gain as a function of her interaction strength $\epsilon$. When $\epsilon = 0$, Eve guesses the classical bit string without measurement, while for $\epsilon =1$, Eve strongly interacts with Alice's state. Right: Alice and Bob's QBER as a function of $\epsilon$.}\label{fig:weak measure}
\end{figure}

After laying the foundation for the analysis of variable-strength measurements with and without ancillary systems, we will now focus our discussion on the case when Eve's weak measurement attack utilizes an ancillary quantum system which is a continuous function. After the mean value of the pointer's function shifts according to Alice's encoded information, Eve performs projective measurements on her pointer and achieves a classical outcome that is correlated with  Alice's original encoded information (where she chooses the same basis as Alice). Eve post-processes her information in the following way: If Eve measures in the conjugate (incompatible) basis, the pointer's mean position does not change and her outcome is equivalent to a guess. 
The following BB84 QKD simulation models an eavesdropper  who performs weak measurements using a Gaussian-pointer apparatus. The results of this attack are plotted in Fig.~\ref{fig:weak measure}, showing Eve’s information gain $G$ and Alice and Bob’s QBER $Q$ as functions of the interaction strength $\epsilon$. In the simulation, Eve's Gaussian pointer shifted according to $\epsilon$
and then her outcome was sampled from the resulting Gaussian function. Then, Eve's pointer was traced-out and Bob performed a (strong) projective measurement. The results of averaging this process for each $\epsilon$ to calculate  $G$ and $Q$ are shown in Fig.~\ref{fig:weak measure} (for each value of $\epsilon$, our simulations were conducted with 10,000 runs). In the following sections, and in particular within  Sec.~\ref{sec:pointer}, Eve uses various types of pointers, corresponding to different continuous  wavefunctions, proving that the pointer's shape also has impact on the information gain-disturbance trade-off.

\subsection{Partial measurements attack}\label{sec:partial}
Partial measurements \cite{elitzur2001nonlocal} may be more practical for Eve to steal a desired fraction from a secret key in a QKD protocol, because she performs standard projective measurements without the requirement of weak interaction with a probe. Partial measurement has been studied in \cite{okamoto2023experimentally} and also enables to extract anomalous weak values. In the attack below, Eve partially measures Alice's qubits leading to the monitoring affine map that was described in Eq.~\eqref{eq: monitoring}. Due to the linearity of the monitoring map, one can easily calculate $Q$ and $G$. Alice and Bob's QBER and Eve's information gain depend linearly on Eve's interaction strength $\epsilon$ and yield the following relations for a BB84 scenario:
    \begin{equation} \label{Eq : G}
   G =  \frac{1}{2} + \frac{\epsilon}{4},
\end{equation}
\begin{equation}\label{eq: Q}
    Q = \frac{\epsilon}{4}.
\end{equation}
This linear connection allows Eve to tune the amount of information that she will steal while controlling Alice and Bob's QBER. In addition, Eve’s pointer is assumed to be noiseless (i.e., the meter introduces no intrinsic noise). However, as discussed in Sec.~\ref{sec:experiment}, this assumption does not necessarily hold in practice. Furthermore, we show that neglecting the nonlinear terms and performing the partial measurement attack is optimal for an eavesdropper targeting QKD protocols based on MUBs.

\section{Comparison between various pointer wavefunctions}\label{sec:pointer}
This section investigates how the characteristics of Eve's measurement device pointer influence the effectiveness of her eavesdropping strategy.
In the case of probe-based weak measurements, the effect on the density matrix results in the monitoring channel defined in Eq.  \eqref{eq: monitoring}. However, the monitoring channel is an approximation that neglects the properties of the pointer’s continuous wavefunction.
In this section, numerical tools were used to plot the exact wavefunction disturbance and Eve’s information gain without omitting the pointer’s shape. To generalize the discussion, let us assume that Alice sends the following density matrix:
\begin{equation}
    \rho = \sum_{i,j\in\{0,1\}}\alpha_{ij}\ket{i}\bra{j}.
\end{equation}
Thus, the contribution from indices $i,j$ is
\[
    \rho_{ij} = \alpha_{ij}\ket{i}\bra{j}.
\]
After weakly coupling Alice’s state to a pointer and performing the variable-strength attack, Eve’s interference modifies Alice’s state. Alice’s local state is then obtained by tracing out Eve’s measurement pointer:
\begin{equation}
    \tilde{\rho} = \sum_{i,j} \chi_{ij}\,\rho_{ij}.
\end{equation}
Here, $\chi_{ij}$ is the response function acting on the encoded state, defined as
\begin{equation}\label{eq:coupling}
    \chi_{ij} = \braket{\phi(x-\epsilon\lambda_i)\mid \phi(x-\epsilon\lambda_j)},
\end{equation}
where $\ket{\phi(x)}$ is the pointer’s wavefunction defined in Sec.~\ref{Sec_weak_measuremet}.

Using numerical tools, we will calculate $Q/G$ for a variety of continuous wavefunctions that may fit the pointers' PDF. Equation \eqref{eq:coupling} represents a convolution between the shifted pointer for each eigenvalue and its counterpart. It is evident that the diagonal terms preserve the unit norm of $\rho$ since $\chi_{ii} = 1$ for all $i$. However, the farther the pointer is shifted, the more information is lost from the off-diagonal terms $\chi_{ij}$, where $i \neq j$. The first pointer PDF we test is a Gaussian, defined as:
\begin{equation}
    \phi(x) = \frac{1}{(2\pi \Delta^2)^{\frac{1}{4}}}e^{\frac{-x^2}{4\Delta^2}}.
\end{equation}
With the pointer wavefunction written as $|\phi\rangle=\int \phi(x)\,|x\rangle\,dx$, the Gaussian-pointer eavesdropping scenario is analyzed in Sec.~\ref{Sec:weak_measurement_attack}, and the corresponding $G$ and $Q$ are plotted in Fig.~\ref{fig:weak measure}.
Other functions that Eve may use are the ``rect'' and ``triangle'', whose derivatives are not continuous. These pointer functions will help us to understand the relation between the pointer's shape and the $Q$, $G$ parameters.

\begin{figure}[h]
    \centering
    \includegraphics[width=0.5\linewidth]{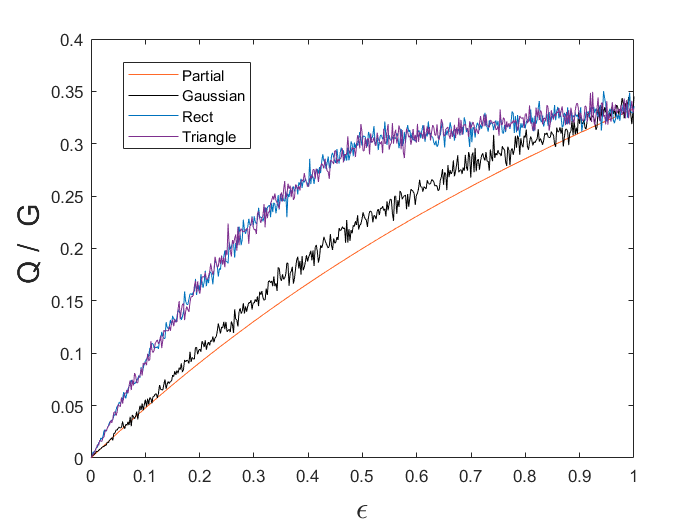}
    \caption{The ratio between disturbance and information gain, $\frac{Q}{G}$, as a function of the interaction strength $\epsilon$ is plotted for probeless partial measurements and for probe-based variable-strength measurements with a Gaussian pointer, rectangular pointer, and triangular pointer. Among these, partial measurement seems the most efficient due to its definite, sharp outcomes. The Gaussian pointer appears to be more efficient than the rectangular and triangular pointers.}
    \label{Fig:delta_q_I}
\end{figure}

The rectangular pointer wavefunction is defined as:
\begin{equation}
r(x)
\;=\;
\frac{1}{L}\,\mathrm{rect}\!\Bigl(\frac{x - x_0}{L}\Bigr)
\;=\;
\begin{cases}
\dfrac{1}{L}, & \lvert x - x_0\rvert < \dfrac{L}{2},\\[8pt]
0, & \text{otherwise}.
\end{cases}
\end{equation}
And the triangular pointer wavefunction is:
\[
t(x)
= \frac{1}{L}\,\mathrm{tri}\!\Bigl(\frac{x - x_0}{L}\Bigr)
=
\begin{cases}
\dfrac{L - \lvert x - x_0\rvert}{L^2}, 
& \lvert x - x_0\rvert < L,\\[6pt]
0, 
& \text{otherwise}.
\end{cases}
\]
When the pointer is initialized at $x_0 = 0$, and the eigenvalues of the measured observable are $\lambda_1= 1, \lambda_2= -1$. The results of Eq.~\eqref{eq:coupling} for each pointer are presented in Fig.~\ref{Fig:pointers}.
\begin{figure}\label{Fig:pointers}
    \centering
    \includegraphics[width=0.5\linewidth]{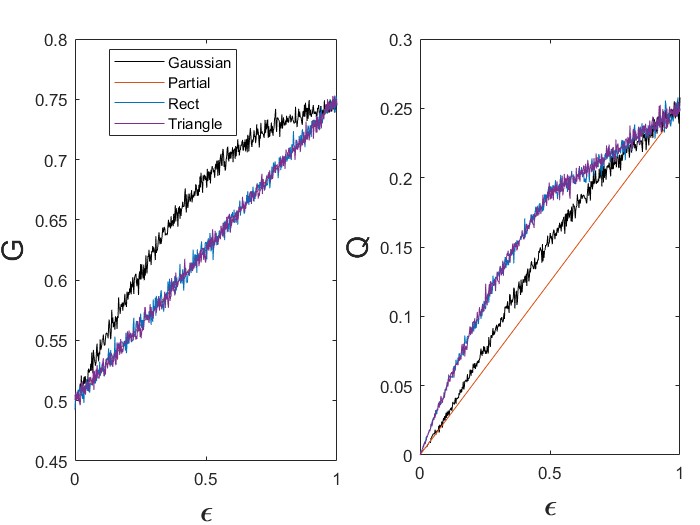}
    \caption{$G$ and $Q$ as functions of the interaction strength, $\epsilon$, are plotted for the Gaussian, partial measurement, rectangular pointer, and triangular pointer. It can be observed that the rectangular and triangular pointers lead to similar performance, whereas the Gaussian pointer is more efficient.}
    \label{Fig:pointers}
\end{figure}
It can be seen that the partial measurement is linear in $G$ and $Q$.
Numerical simulations suggest that the ``rect'' and ``triangle'' pointer shapes are suboptimal for eavesdropping, likely due to the discontinuities in their derivatives, which hinder accurate information extraction.
However, it is not yet clear whether this is superior to the Gaussian pointer. To select the preferable attack for Eve, the efficiency metric $Q/G$ was evaluated, balancing information gain against Bob’s QBER; the strategy minimizing $Q/G$ is deemed most efficient.
According to this metric, the results, shown in Fig.~\ref{Fig:delta_q_I}, indicate that partial measurement is the preferred method for eavesdropping within this model. However, this may not hold for other applications, particularly those involving sequential weak measurements.

\section{Photon number splitting attack combined with partial measurements}\label{PNS and Weak}
\begin{figure}[t]
    \centering
    \begin{subfigure}[t]{0.48\textwidth}
        \centering
        \includegraphics[scale=0.3]{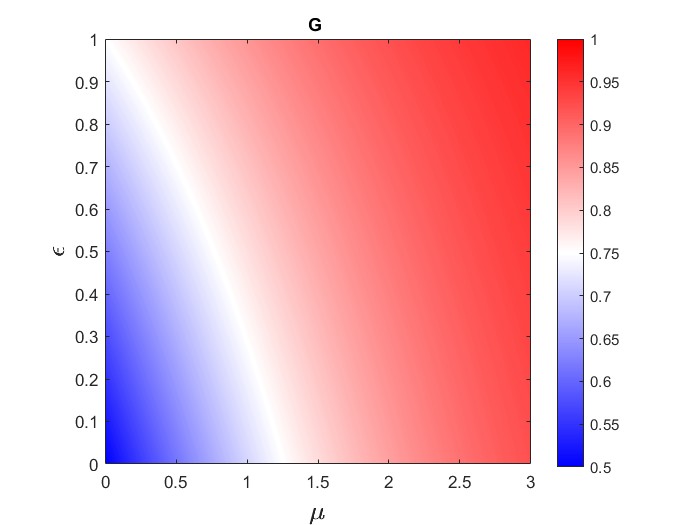}
        \caption{Information gain $G$ as a function of $\mu$ and $\epsilon$}
        \label{fig:gain}
    \end{subfigure}
    \hfill
    \begin{subfigure}[t]{0.48\textwidth}
        \centering
        \includegraphics[scale=0.3]{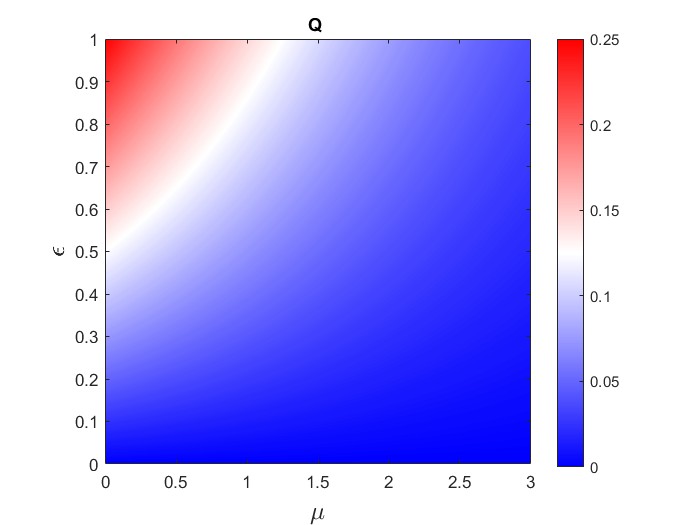}
        \caption{QBER $Q$ as a function of $\mu$ and $\epsilon$}
        \label{fig:qber}
    \end{subfigure}
    \caption{Heatmaps of the information gain and QBER over the photon rate $\mu$ and interaction strength $\epsilon$.}
    \label{fig:gain_qber}
\end{figure}

To complete the theoretical discussion of Eve’s potential attacks, a PNS attack \cite{huttner1995quantum} is analyzed in combination with subsequent partial measurements. Using the PNS attack, Eve splits multi-photon pulses resulting from WCS into a single photon which she can measure and the remaining photon(s) which will be transmitted to Bob. Combining the PNS attack with partial measurements may enhance Eve's flexibility and allow her to choose the optimal trade-off between QBER and information gain. The proposed protocol is therefore as follows. Initially, Eve performs the standard PNS attack, meaning that if the pulse contains more than one photon, Eve gains ``free'' information from the pulse she split by means of a projective measurement (assuming Eve has quantum memory). However, if the pulse is a single-photon pulse, she will use the aforementioned partial measurement method. A straightforward probabilistic argument then shows that Eve's gain (per detected non-vacuum pulse) $G$ will be given as: 
\begin{equation}\label{Eq:Eve gain co}
    G = \frac{P(n=1)(\frac{1}{2} + \frac{\epsilon}{4}) + P(n>1)}{P(n>0)}. 
\end{equation}
The QBER $Q$ will be:
\begin{equation}\label{Eq: q cow}
    Q = \frac{\epsilon P(n=1)}{4P(n>0)},
\end{equation}
where the probabilities are calculated using the WCS defined in Eq.~(\ref{eq:WCS}).
Thus, in the limit of $\mu \rightarrow 0$, Eqs.~(\ref{Eq:Eve gain co}),(\ref{Eq: q cow}) become Eqs.~(\ref{Eq : G}), (\ref{eq: Q}), illustrated in Fig.~\ref{fig:gain_qber}. However, $\mu \gg 1$ leads to  $Q \approx 0$ and $G \approx 1$. This combined attack shares the limitations of the individual partial measurement and PNS attacks, but without proper countermeasures (such as decoy states \cite{branciard2005security, hwang2003quantum}), Eve's threats become more significant, especially for high-$\mu$ WCS.


\section{Experiment}\label{sec:experiment}
To test the above, we have performed an experiment based on the simplified BB84 protocol~\cite{rusca2018security,boaron2020long}.
In this variant, the sifted key is distilled only from the (time)  Z basis $\{\ket{e},\ket{l}\}$, while the (phase) X basis $\{\ket{+},\ket{-}\}$ is used solely to estimate the phase-error (monitoring) rate, i.e. outcomes obtained in $X$ are discarded for key generation. In our implementation, the optical paths and detectors of the $Z$ and $X$ measurements are not perfectly symmetric (different noise and visibility budgets), so we report both $Q_Z$ and $Q_X$. $Q_Z$ characterizes the data line used for key distillation, whereas $Q_X$ quantifies the monitoring line that bounds the phase error in the security analysis. This experiment was performed using time-bin-encoding in a fiber-coupled setup, suitable for short-distance QKD in data centers. Additional details about the experimental setup appear in Appendix~\ref{appendix : experiment-d} and Fig.~\ref{fig:experimental_illustration}. The chosen implementation is closely related to the approaches in Refs.~\cite{rusca2018security,boaron2020long}. Throughout the experiment, Eve performed partial measurements to extract information from the secret key shared between Alice and Bob, as discussed in Sec.~\ref{sec:partial}. We conducted multiple experiments, each examining different aspects of the protocol. The first was a calibration experiment designed to test the intrinsic attributes of the QKD protocol. In a subsequent set of experiments, Eve implemented an intercept-and-resend attack using tunable partial measurements. While such an attack can be realized using a variable beamsplitter (BS), in our setup, Eve directly measured a fraction of Alice's incoming pulses, with the fraction determined by her interaction parameter. In each experiment, Eve used a different interaction parameter -- from no interaction at all (Eve guesses Alice's classical bits) to strong interaction (standard projective measurement). We finally simulated a more advanced attack in which Eve additionally injected noise into the system, only during Alice and Bob's calibration stage, in an attempt to deceive them and gain further information without being noticed.
\begin{figure}[h]
    \centering
    \includegraphics[width=0.6\linewidth]{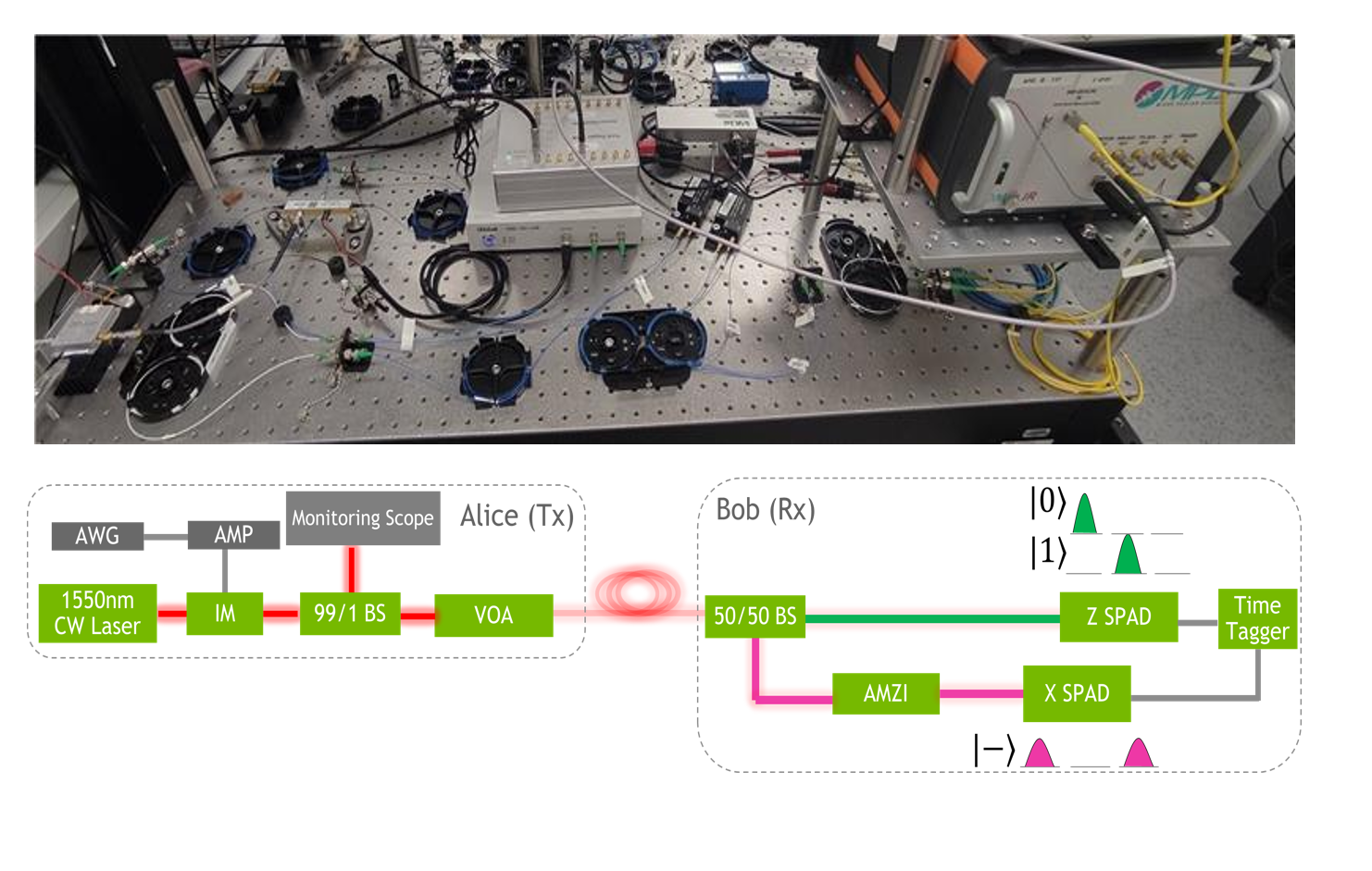}
    \caption{A continuous-wave laser at 
    1550\,nm first passes through an intensity modulator and a variable optical attenuator on the encoder side. At the decoder, a 50:50 beamsplitter (BS) passively selects the measurement basis: one arm goes directly to a single-photon avalanche diode (SPAD) for the time (Z) basis, while the other enters an AMZI to enable interference measurements in the phase (X) basis before reaching its SPAD. All detection events are recorded by a time-tagger.}
    \label{fig:experimental_illustration}
\end{figure}

\begin{figure}[h]
    \centering
    \includegraphics[width=1\linewidth]{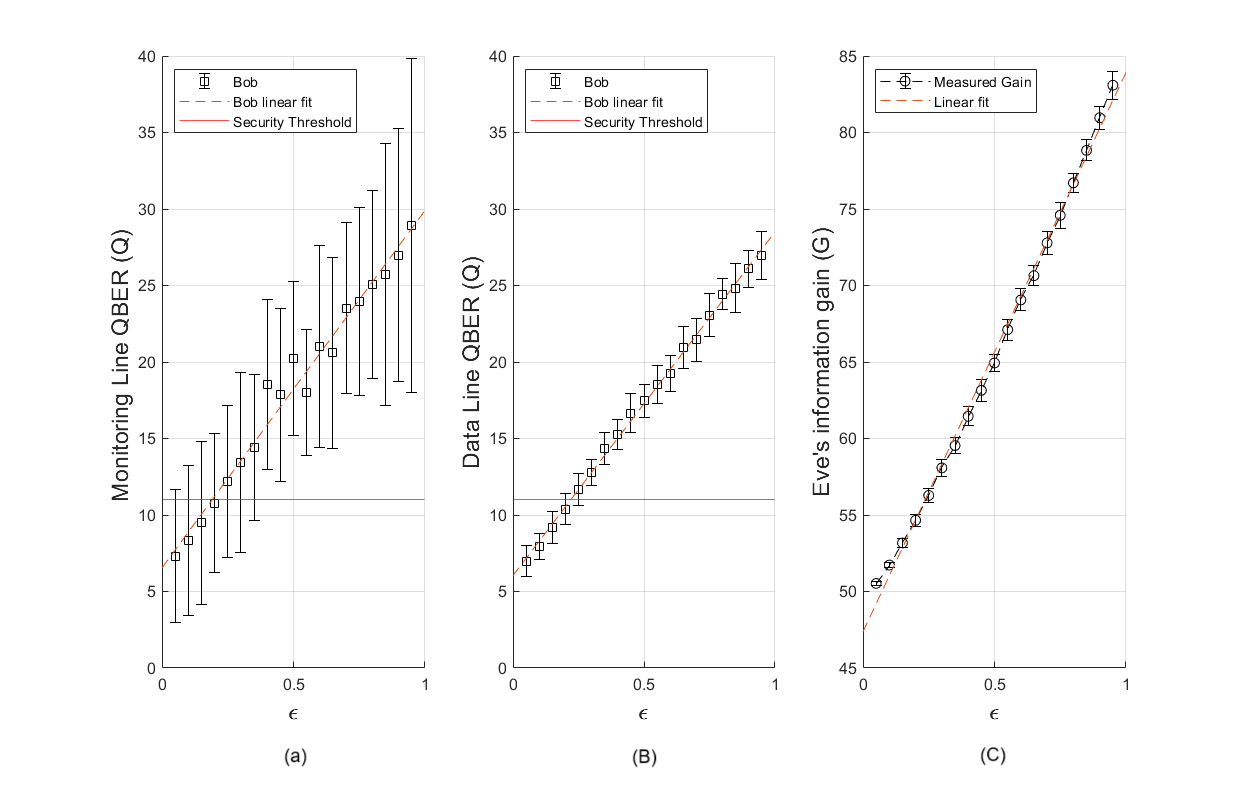}
    \caption{Experimental results for Alice and Bob's QBER (Q) and Eve's information gain (G) are presented. The red line indicates the $11\%$ QBER threshold, at which Alice and Bob abort the protocol. The QBER in the phase basis exhibits larger variance due to the AMZI loss of coherence over time, compared to the time-bin QBER, which is more precise. The information gain does not reach the maximum value due to inherent system noise. The curves are indeed linear, in accordance with the theoretical analysis underlying partial measurements. } 
    \label{fig:experiment}
\end{figure}

The experimental setup included inherent noise leading to a QBER of around 7$\%$. This does not provide Eve with much freedom because of the well-known $\sim 11\%$ QBER threshold \cite{metger2023security, shor2000simple}. The first step examines Eve's attack on the Z basis. Even though in the standard simplified BB84 protocol Alice and Bob evaluate the QBER only in the X basis (monitoring line), both the data line and the monitoring line were partially measured by Eve. The dependence of \(G\) and \(Q\) on the interaction strength $\epsilon$ is shown in Fig.~\ref{fig:experiment}. The theoretical linear relationship remains visible in the experimental data despite inherent losses. The relation derived in Sec.~\ref{sec:QaG} still holds, but it is somewhat distorted due to experimental limitations. Under Eve’s partial measurement attack, she fails to reach the upper bound of her gain parameter \(G\), whereas the parameter \(Q\) is amplified by experimental noise.

 In our experiment, the monitoring line QBER exhibits greater variance than the data line QBER owing to the AMZI, which relies on destructive interference and hence leads to fewer events (and limited statistics). In contrast, the data line employs a simpler measurement process with lower associated losses. Additionally, Fig.~\ref{fig:experiment}(c) depicts the role of internal errors which constrain Eve’s gain parameter \(G\), causing it to fall below its theoretical upper bound.

In this section, we demonstrate Eve’s limited ability to extract a fraction of the sifted key; however, she cannot obtain enough information to compromise the security of the QKD protocol. Next, Eve’s attack on the experimental setup will include an additional noise‐injection step, in which she perturbs system components. This extra loss should mislead Alice and Bob into attributing the degradation to faulty hardware rather than to an external eavesdropper.

\section{Noise injection attack}\label{sec:noise injection}
Relying on our previous analysis of Eve’s partial‐measurement attack on the experimental setup (Sec.~\ref{sec:experiment}),  this section examines the scenario where Eve injects noise into Alice and Bob’s system during the calibration stage by locally intervening with their hardware. Eve artificially increases the noise within the setup (that is, in the monitoring line corresponding to the phase basis) and makes Alice and Bob believe that their QBER is higher than it actually is. Once the calibration stage is over and the QKD protocol begins, Eve refrains from injecting noise and instead performs variable-strength measurements on the pulses, using a tunable interaction strength that corresponds to the noise previously introduced (so that the applied channel approximately takes the same form). The total noise in the system is denoted by $Q_T = Q_{\text{env}} + Q_E$, where $Q_{\text{env}}$ represents the inherent QBER due to environmental factors which also affect Eve and $Q_E$ denotes the noise intentionally introduced by Eve during calibration, which is later translated into the noise induced by her interaction with the qubits.
Thus, Eve's attack might be less detectable for Alice and Bob. In general, her motivation is to make the attack effects seem more natural, thus misleading Alice and Bob regarding the true source of the QBER. The experimental results demonstrate that such an attack is indeed feasible. However, as we shall see, it still does not allow Eve to obtain a significant portion of the secret key.
\begin{figure}[h]
    \centering
    \includegraphics[width=0.6\linewidth]{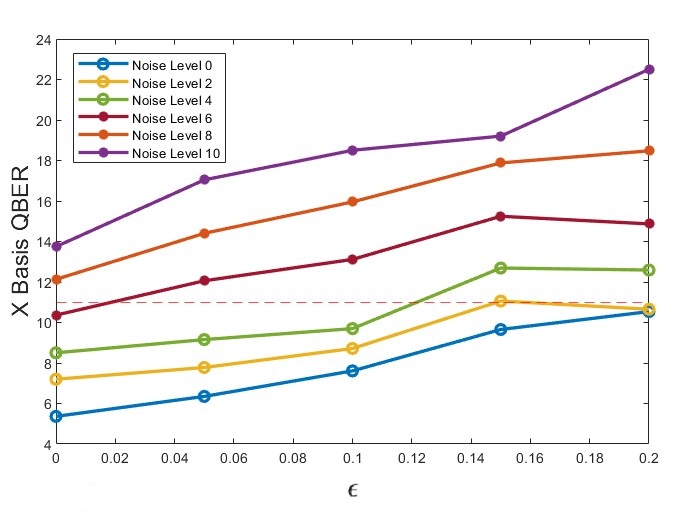}
    \caption{The QBER at the monitoring line is analyzed as a function of Eve's measurement strength, denoted by $\epsilon \in [0, 0.2]$. Each curve corresponds to the voltage applied by Eve to Bob's AMZI as part of the monitoring line at the calibration stage of the protocol. The applied voltage is divided into 10 incremental steps of 20\,mV each (where Bob's standard working voltage is 0.92\,V). The artificially increased voltage lowers the interference visibility and hence amounts to greater noise injection into the system, which results in higher QBER.}
    \label{fig:noise injection}
\end{figure}
The noise injection is bucketed into discrete levels corresponding to the magnitude of voltage shifts that Eve applies to Bob's AMZI. This process induces statistical noise characterized by an average value, as depicted in Fig.~\ref{fig:noise injection}. When $\epsilon = 0$, the total noise reduces to $Q_T = Q_{\text{env}}$; otherwise, $Q_E$ increases the QBER beyond the environmental contribution.

\section{Conclusion}\label{sec:conco}
We investigated information–disturbance relations for variable-strength eavesdropping strategies, contrasting probe-based weak measurements with probeless (partial) measurements and applying the analysis to time-bin-encoded BB84. Within a two-MUB setting, the comparison shows that optimized partial measurements attain a larger information gain for a given induced disturbance than their probe-based counterparts. We then instantiated these ideas in a fiber-coupled, time-bin implementation of simplified BB84 with weak coherent states, and we observed the expected near-linear relation between Eve’s interaction strength, Alice–Bob’s QBER, and Eve’s information gain. We further examined composite threats, i.e., a PNS-assisted partial-measurement strategy and a calibration-stage noise-injection tactic that can partially mask Eve’s later interaction.

Our results reinforce that, even under these variable-strength attacks, the protocol remains secure when standard abort thresholds and parameter estimation are enforced. Eve’s net information advantage is constrained by the disturbance she inevitably introduces. At the same time, our analysis and experiments clarify which countermeasures are most impactful in such cases: (i) decoy-state methods to close PNS avenues; (ii) stronger parameter estimation via additional test bases (e.g., sporadic X-basis checks on the data line or six-state variants); (iii) calibration hardening with authenticated procedures and continuous monitoring; and (iv) optical isolation, input-power monitors, and detector-gate randomization to reduce stealth channels.

Overall, the results provide an applied, side-by-side assessment of probeless and probe-based variable-strength measurements, supported by fiber-link data and by practically relevant composite attacks, and they translate directly into deployable defenses and concrete avenues for future work.

\section*{Acknowledgements} 
This work was supported by the Israel Innovation Authority under the Quantum Communication Consortium, by the European Union’s Horizon Europe research and innovation programme under grant agreement No. 101178170 and by the Israel Science Foundation under grant agreement No. 2208/24.

\bibliographystyle{ieeetr}
\bibliography{references.bib}
\appendix

\section{Experimental details}\label{appendix : experiment-d}
 In the experiment we use a 1550\,nm continuous wave laser (SanTec TSL-570), and a LiNbO$_3$ intensity modulator (iXblue MXER-LN-10), employing an RF signal originating from an arbitrary wave generator (AWG, Keysight M8196A) and amplified by an RF amplifier (iXblue DR-VE-10-MO) to carve optical pulses out of the laser beam. Then the signal passes through a 99:1 fiber BS which reflects 99\% of the optical power to a monitoring oscilloscope (Keysight N1000A, 86105D). The remaining 1\% is attenuated via a variable optical attenuator (VOA, Oz Optics DD-600-MC), which is then transmitted to Bob via a short optical fiber, typical of data centers. On the receiver side the qubit enters a $50/50$ fiber BS. In the first outgoing port the pulse reaches the single-photon avalanche diode (SPAD, MPD PDM-IR) and in the second port the pulse passes through the AMZI (Kylia 06272-MINT-IN PM-1.19GHz-L-FCUPC), later measured by a similar SPAD. Furthermore, there is a time tagger (Swabian Instruments Time Tagger Ultra) connected to both SPADs and AWG. In order to implement the partial measurements, Eve needs to place a BS between Alice and Bob. However, due to hardware constraints, the eavesdropping was implemented with time-division multiplexing (TDM). Denoting the total duration of Alice's transmission by $T$, and Eve's measurement duration by $t$, the proportion between Eve's measurement window and Alice's transmission time is $R = \frac{t}{T}$, where $ 0 \leq R \leq 1$ is proportional to the interaction strength $\epsilon$.

\end{document}